\documentclass{PoS}

\usepackage[fleqn]{amsmath}

\newcommand{\be}{\begin{equation}}\newcommand{\ee}{\end{equation}}
\newcommand{\bea}{\begin{eqnarray}}\newcommand{\eea}{\end{eqnarray}}

\newcommand{\nn}{\nonumber}
\renewcommand{\thefootnote}{\@fnsymbol\c@footnote}
\renewcommand{\[}{\langle\!\langle}\renewcommand{\]}{\rangle\!\rangle}

\title{Heavy quarkonium spectroscopy in pNRQCD\\ with lattice QCD input}

\ShortTitle{Heavy quarkonium spectroscopy in pNRQCD with lattice QCD input}

\author{\speaker{Yoshiaki Koma}%
\thanks{Y.K. is partially supported by the Ministry of Education, Science, 
Sports and Culture, Japan, Grant-in-Aid for Young Scientists (B) (24740176).}\\
Numazu College of Technology\\
E-mail: \email{koma@numazu-ct.ac.jp}}

\author{Miho Koma%
\thanks{M.K. is supported by Japan Society for the Promotion 
of Science (JSPS), Grant-in-Aid for JSPS Fellows (20$\cdot$40152).}\\
Numazu College of Technology\\
E-mail: \email{m-koma@numazu-ct.ac.jp}}

\abstract{The charmonium and bottomonium mass spectra are
investigated  in potential nonrelativistic QCD (pNRQCD) with the 
heavy quark potential computed by lattice QCD simulations.
The potential consists of a static potential and relativistic corrections
classified in powers of the inverse of heavy quark mass~$m$, and
the effects of the $O(1/m)$ and $O(1/m^{2})$ spin-orbit corrections on
the mass spectra are examined systematically.
The pattern of the mass spectra is found to be in fairly good agreement with experimental data,
in which the $O(1/m)$ correction gives an important contribution.
}

\FullConference{The 30th International Symposium on Lattice Field Theory\\
June 24 - 29,  2012\\
Cairns, Australia}

\begin{document}

\section{Introduction}

\par
It is of great  challenge to understand a systematic pattern of heavy
quarkonium mass spectra~\cite{Beringer:1900zz}
from nonperturbative QCD.
Lattice QCD simulations offer strong tools to gain an understanding of nonperturbative QCD,
which enable us to compute the mass spectra of quarkonia
directly by evaluating appropriate correlation functions of operators on the lattice.
However, 
due to a hierarchical mixture of ultraviolet and infrared physics in the quarkonium system,
it is not straightforward to deduce the underlying mechanism 
responsible for the mass spectra from the conventional simulation results.

\par
One way to solve the hierarchical problem in the quarkonium system is
to employ effective field theories (EFTs). In this context, the use of 
nonrelativistic QCD (NRQCD)~\cite{Caswell:1985ui,Bodwin:1994jh} and 
potential NRQCD (pNRQCD)~\cite{Brambilla:2004jw,Brambilla:2000gk,Pineda:2000sz} 
have been proposed.
The aim of EFTs is generally to have better control of a hierarchy
of energy scales in an original theory for a high precision calculation. 
For instance, suppose there exist two typical energy scales which satisfy 
$E_{\rm high} \gg E_{\rm low}$ in the original theory,
then by integrating the scale above $E_{\rm high}$, 
an EFT with the energy scale $E_{\rm low}$ can be derived.
The high energy contribution is 
to be incorporated in the effective couplings 
of the interactions in the EFT,
which are referred to as the matching coefficients,
at any desired fixed order of perturbation theory 
in the small ratio $E_{\rm low}/E_{\rm high}\ll 1$.

\par
In the quarkonium system, it is considered that there exists a hierarchy of three types of energy scales,
$m \gg  mv \gg mv^{2}$,
with a heavy quark mass $m \gg \Lambda_{\rm QCD}$ and a quark velocity $v$.
NRQCD has been derived by integrating the energy scale above $m$ in QCD,
and pNRQCD by integrating further the energy scale above $mv$ in NRQCD.
Nonrelativistic nature of the quarkonium system then becomes manifest in these EFTs, 
and notably, a quantum mechanics-like hamiltonian emerges in pNRQCD.
The matching coefficients in pNRQCD are dependent on the distance between 
a heavy quark and antiquark and eventually classified in powers of $1/m$.
Thus a set of these matching coefficients 
can be regarded as {\em the heavy quark potential}
consisting of a static potential and relativistic corrections.
Therefore, once these matching coefficients are determined,
various properties of quarkonia, such as the mass spectra, can be investigated systematically 
by solving the Schr\"odinger equation.

\par
The nonperturbative matching is crucial for obtaining the heavy quark potential in pNRQCD
as the energy scale of $mv$ can be the same order of the magnitude as $\Lambda_{\rm QCD}$.
For this purpose, the present authors have been performing lattice QCD simulations 
and obtained accurate data so far up to the $O(1/m^{2})$ 
corrections~\cite{Koma:2005nq,Koma:2006si,Koma:2006fw,Koma:2009ws,Koma:2010zz}.
The aim of the present report is then to demonstrate a spectroscopy analysis of 
quakonia in pNRQCD with the lattice QCD results;
the mass spectra of charmonium and bottomonium are computed by
solving the Schr\"odingier equation, and the effect of relativistic corrections,
in particular, the $O(1/m)$ and the $O(1/m^{2})$ spin-orbit corrections
are examined systematically.
The pattern of the mass spectra is found to be in fairly good agreement with experimental data,
in which the $O(1/m)$ correction gives an important contribution.

\section{pNRQCD and the heavy quark potential}

\par
We begin by describing 
how the hierarchy of energy scales in the heavy quarkonium system, $m \gg  mv \gg mv^{2}$, 
is controlled along the derivation of NRQCD and pNRQCD and how the 
heavy quark potential is defined nonperturbatively.

\par
The first step is to integrate the energy scale above $m$ in QCD, which leads to NRQCD.
The meaning of the integration is that an interaction of heavy quarks via gluons 
whose momenta are larger than $m$ is no more visible as a dynamical one
and a creation or an annihilation of a heavy quark and antiquark pair 
cannot be seen within the resolution of the resulting obscured vacuum.
Then, the heavy quark and antiquark are treated as independent external fields, 
respectively.

\par
Let $\psi$ and  $\chi_{c}$ be a quark and an antiquark annihilation operators, respectively,
the NRQCD Lagrangian density  with the leading interaction terms of the heavy quark and antiquark
is written as~\cite{Caswell:1985ui,Bodwin:1994jh}
\bea
&&{\cal L}_{\rm NRQCD}
=
\psi^{\dagger} (iD_{0} +\frac{\pmb{D}^{2}}{2m}+\frac{\pmb{D}^{4}}{8m^{3}})\psi +
(\psi \to \chi_{c}) -\frac{1}{4}F_{\mu\nu}F^{\mu\nu}\nn\\
&&
+
\psi^{\dagger} ({c_{F}}g\frac{\pmb{\sigma}\!\cdot\!\pmb{B}}{2m}
\!+\!{c_{D}}g\frac{\pmb{D}\! \cdot\! \pmb{E}\! -\!\pmb{E}\! \cdot\! \pmb{D}}{8m^{2}}
\!+\!i{c_{S}} g \frac{\pmb{\sigma}\!\cdot \! (\pmb{D}\!\!\times\!\!\pmb{E}
\!-\!\pmb{E}\!\!\times\!\!\pmb{D})}{8m^{2}}
)\psi
+(\psi \to \chi_{c})
\nn\\
&&
+\frac{d_{ss}}{m^{2}}\psi^{\dagger}\!\psi\chi_{c}^{\dagger}\!\chi_{c}
+
\frac{d_{sv}}{m^{2}}\psi^{\dagger}\!\pmb{\sigma}\psi\!\cdot\!\chi_{c}^{\dagger}\!\pmb{\sigma}\chi_{c}
+
\frac{d_{vs}}{m^{2}}\psi^{\dagger}T^{a}\psi \chi_{c}^{\dagger}T^{a}\chi_{c}
+
\frac{d_{vv}}{m^{2}}\psi^{\dagger}T^{a}\pmb{\sigma}\psi\!\cdot\!\chi_{c}^{\dagger}\!T^{a}\pmb{\sigma}\chi_{c}
\;,
\eea
where  $\pmb{B}$ and $\pmb{E}$ represent the color-magnetic and color-electric fields, and 
$D_{0}$ and $\pmb{D}$ are the time and spatial components of covariant derivative.
The Pauli matrix  is denoted as $\pmb{\sigma}$, and $T^{a}$~$(a=1,...,8)$ are the SU(3) color generators.
The factors $c_{F}$, $c_{D}$, $c_{S}$, $d_{ss}$,  $d_{sv}$,  $d_{vs}$,  $d_{vv}$
are the matching coefficients, which are
to be determined so as to reproduce the same quantity, such as scattering
amplitudes, both in QCD and NRQCD 
at any desired fixed order in perturbation theory of~$\alpha_{s}=g^{2}/(4\pi)$.
This matching is always performed perturbatively as $m \gg \Lambda_{\rm QCD}$.
In this way, the high energy momenta of gluons are 
integrated into the matching coefficients,
while the remaining gluons in NRQCD carry nonrelativistic low energy momenta.
For instance, if the matching scale is chosen just to be~$m$, 
the matching coefficients of the bilinear term have the forms,
$c_{F}=1+ \frac{\alpha_{s}}{2\pi}\left (C_{F}+C_{A} \right )+ O(\alpha_{s}^{2})$,
$c_{D}=1+ \frac{\alpha_{s}}{2\pi} C_{A}+ O(\alpha_{s}^{2})$,
$c_{S}=2c_{F}-1$~\cite{Manohar:1997qy}, 
where $C_{F}=4/3$ and $C_{A}=3$ are the eigenvalues of the quadratic 
Casimir operators of the fundamental and adjoint representation in SU(3), respectively,
and those of the four-Fermi contact terms are 
$d_{ss}=\frac{2}{3}\pi \alpha_{s}+O(\alpha_{s}^{2})$,
$d_{sv}=-\frac{2}{9}\pi \alpha_{s}+O(\alpha_{s}^{2})$, 
$d_{vs}=-\frac{1}{2}\pi \alpha_{s}+O(\alpha_{s}^{2})$, 
$d_{vv}=\frac{1}{6}\pi \alpha_{s}+O(\alpha_{s}^{2})$~\cite{Pineda:1998kj}.
More general forms of the  matching coefficients are provided in 
the original references~\cite{Manohar:1997qy,Pineda:1998kj}.

\par
The second step is to integrate the energy scale above $mv$ in NRQCD, which leads to pNRQCD.
A merit of further integration is to avoid 
a complication due to a mixture of the remaining two energy 
scales $mv$ and $mv^{2}$ in NRQCD, which may affect a power
counting of operators.
The procedure is similar to the first step, i.e. 
one computes the same quantity in NRQCD and pNRQCD.
However, in contrast to the first step, this matching must be performed nonperturbatively,
since $mv$ can be the same order of magnitude as $\Lambda_{\rm QCD}$.
The procedure proposed by Brambilla {\it et al.}~\cite{Brambilla:2004jw,Brambilla:2000gk} 
for the nonperturbative matching is called the quantum mechanical matching, 
which consists of the following steps.
Firstly one writes the NRQCD hamiltonian with the $1/m$ expansion,
$H_{\rm NRQCD} =H^{(0)}+\frac{1}{m}H^{(1)}+\frac{1}{m^{2}}H^{(2)}+\cdots$
and then evaluates the expectation values of $H^{(i\ge 1)}$
with an eigenstate of $H^{(0)}$ projected onto a color-singlet $q$-$\bar{q}$ state.
Typically, the expectation values of $H^{(i\ge 1)}$ are represented with  
color-field strength correlators on the $q$-$\bar{q}$ source.
Then, the ground state energies of these expectation values are compared 
to the pNRQCD matching coefficients, 
which are also classified in powers of~$1/m$,
\be
V_{\rm pNRQCD}=V^{(0)}+\frac{1}{m}V^{(1)}+\frac{1}{m^{2}}V^{(2)}+\cdots \;,
\label{eq:pnrqcd-pot}
\ee
where $V^{(i\geq 0)}$ are local in time but nonlocal in space 
(dependent on the $q$-$\bar{q}$ distance $r$), hence they can be regarded as the components of 
the heavy quark potential.
$V^{(0)}$ is expressed with the Wilson loop, and thus, related to the static potential.
$V^{(i\geq 1)}$ are clearly due to finiteness of the quark mass, which are
expressed with the color-field strength correlators and identified as the relativistic corrections to the 
static potential.
So far these expressions are provided up to of $O(1/m^{2})$~\cite{Brambilla:2000gk,Pineda:2000sz}.
The second term in Eq.~\eqref{eq:pnrqcd-pot}, corresponding to the $O(1/m)$ correction, 
is expressed as~\cite{Brambilla:2000gk}
\be
V^{(1)}(r)=  -\!\int\!\! dt\, t\[\pmb{E}(0)\!\cdot\! \pmb{E}(0)\] \; .
\label{eq:o1m}
\ee
This term is typical in QCD due to  the self interaction of gluons (three-gluon vertex).
The double bracket $\[ {\cal O} \]$ indicates that an expectation value of an operator
 ${\cal O}$ on a $q$-$\bar{q}$ source  
normalized by the expectation value of the  $q$-$\bar{q}$ source itself.
The third term contains the spin-dependent and spin-independent corrections.
The spin-orbit correction in $V^{(2)}$, which is relevant in the following analysis,
 is expressed as~\cite{Pineda:2000sz}
\be
V_{ls}^{(2)}(r) =
\Biggl ( \frac{c_{S}}{2r}\frac{dV^{(0)}}{dr}\!+\!
\frac{c_{F}}{r}(
\underbrace{\epsilon_{ijk}\hat{\pmb{r}}_{i}\!\!\int\!\! dt\,  t \[\pmb{B}^{j} (0)\pmb{E}^{k}(0) \]}_{\equiv ~V_{1}'(r)}
+ \underbrace{\epsilon_{ijk}\hat{\pmb{r}}_{i}\!\!\int\!\! dt\, 
 t \[\pmb{B}^{j}(0) \pmb{E}^{k}(r) \]}_{\equiv ~V_{2}'(r)}) \Biggr )
\pmb{l}\cdot \pmb{s}
 \; .
\label{eq:o2ls}
\ee
The arguments of the color-field strength tensors in Eqs.~\eqref{eq:o1m} and~\eqref{eq:o2ls}, 
such as $0$ or $r$,  mean that a color-field strength tensor is attached to either quark or antiquark.
The variable $t$ is a relative temporal distance between the two color-field strength
tensors, which is to be integrated.
The matching coefficients $V^{(i\geq 0)}(r)$ are then computed by lattice QCD simulations.

\section{Brief summary of the lattice QCD results}

\par
The present authors have been studying the heavy quark potential in pNRQCD 
by using lattice QCD simulations within the quenched
approximation~\cite{Koma:2005nq,Koma:2006si,Koma:2006fw,Koma:2009ws,Koma:2010zz}.
The numerical procedure has been to use the Polyakov loop correlation
function for the $q$-$\bar{q}$ source, and compute the field strength correlators
by employing the multilevel algorithm,  and evaluate them with transfer matrix theory.
So far the potential with the $O(1/m^{2})$ relativistic corrections have been computed 
accurately up to the distance $r\simeq 1~{\rm fm}$.

\par
Fig.~\ref{fig:lattice} summarizes a part of the lattice QCD results which are relevant in the following
spectrum analysis, the static potential, the $O(1/m)$ correction, and the $O(1/m^{2})$ spin-orbit correction.
The Sommer scale $r_{0}=0.50$~fm has been used to fix the lattice spacing for each $\beta$ value.
The functional forms of the potentials from short to long distances have been determined 
by the following fitting functions, 
\be
V^{(0)}(r) =-\frac{\alpha}{r}+\sigma r + c^{(0)} \;,
\ee
\be
V^{(1)}(r) =- \frac{9\alpha^{2}}{8r^{2}}+ \sigma^{(1)} \ln r+ c^{(1)}\;,
\ee
\be
V^{(2)}_{ls}(r)  = \left (\frac{{c_{s}}}{2r}\frac{dV^{(0)}}{dr}
+ \frac{{c_{F}}}{r}(V_{1}'+ V_{2}')\right )
\pmb{l}\!\cdot\!  \pmb{s}
\;, \quad
V_{1}' = -(1\!-\!\epsilon)\sigma \;, \quad
V_{2}' =\frac{\alpha}{r^{2}}+\epsilon \sigma \;.
\ee
The fitting parameters have been found to be
\be
\alpha = 0.297\;, \quad  \sigma=1.06~{\rm GeV\!/fm} \;,  \quad
\sigma^{(1)}=0.142~{\rm GeV^2} \;,  \quad\epsilon = 0.2 \; .
\ee
As can be seen in Fig.~\ref{fig:lattice}, 
the fitting curves nicely describe the behavior of the lattice data.
Note that the leading order perturbation theory results in
$V^{(0)}(r) = - \frac{\alpha}{r}$ with $\alpha \equiv C_{F} \alpha_{s}$,
$V^{(1)}(r) =  -\frac{9\alpha^{2}}{8r^{2}}$, and $V_{1}'(r) = 0$, $V_{2}'(r) = \alpha/r^{2}$.
The short distance behavior of $V^{(0)}$, $V^{(1)}$ and $V_{2}'$ may follow these functions,
although the coupling $\alpha$ is different from the bare one.
It can be observed that both $V_{1}'$ and $V_{2}'$ contain long-ranged nonperturbative tails,
which are related to the string tension in $V^{(0)}$ through the Gromes relation
 $dV^{(0)}/dr =V_{2}'  -V_{1}'$~\cite{Gromes:1984ma}.

\newcommand{\xsize}{7.4cm}
\begin{figure}[!t]
\includegraphics[width=\xsize]{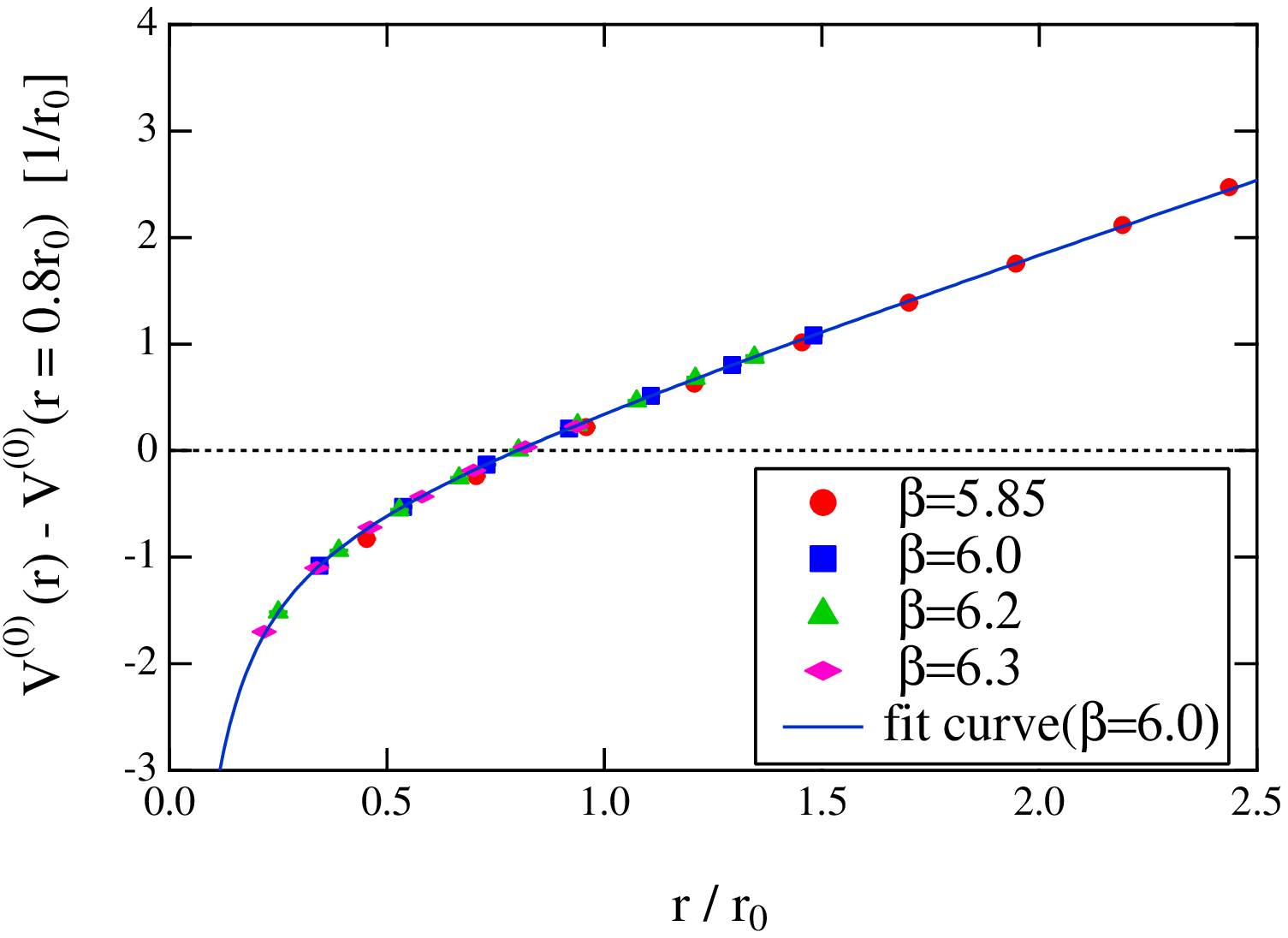}\hspace{0.4cm}%
\includegraphics[width=\xsize]{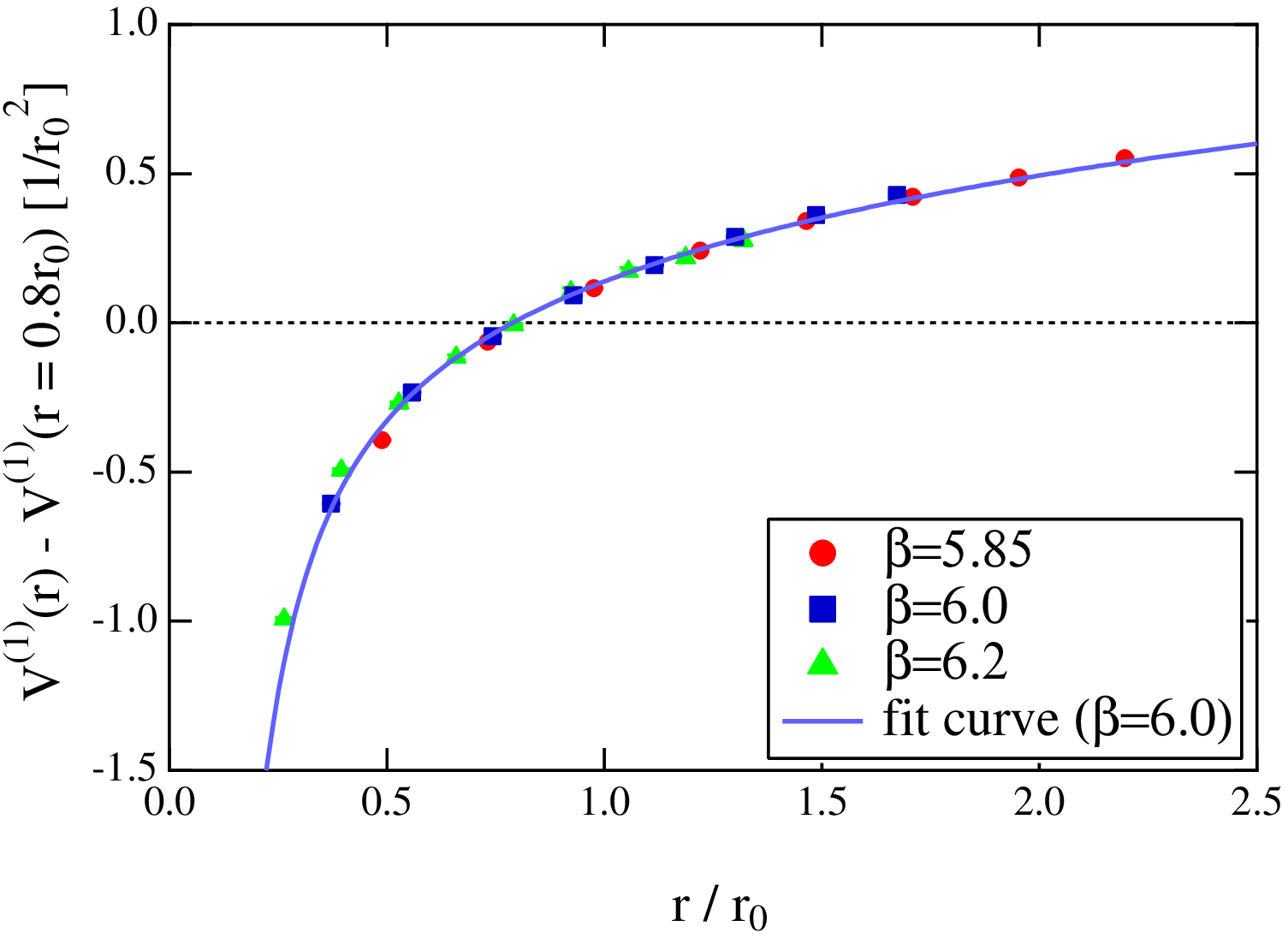}\\[0cm]
\includegraphics[width=\xsize]{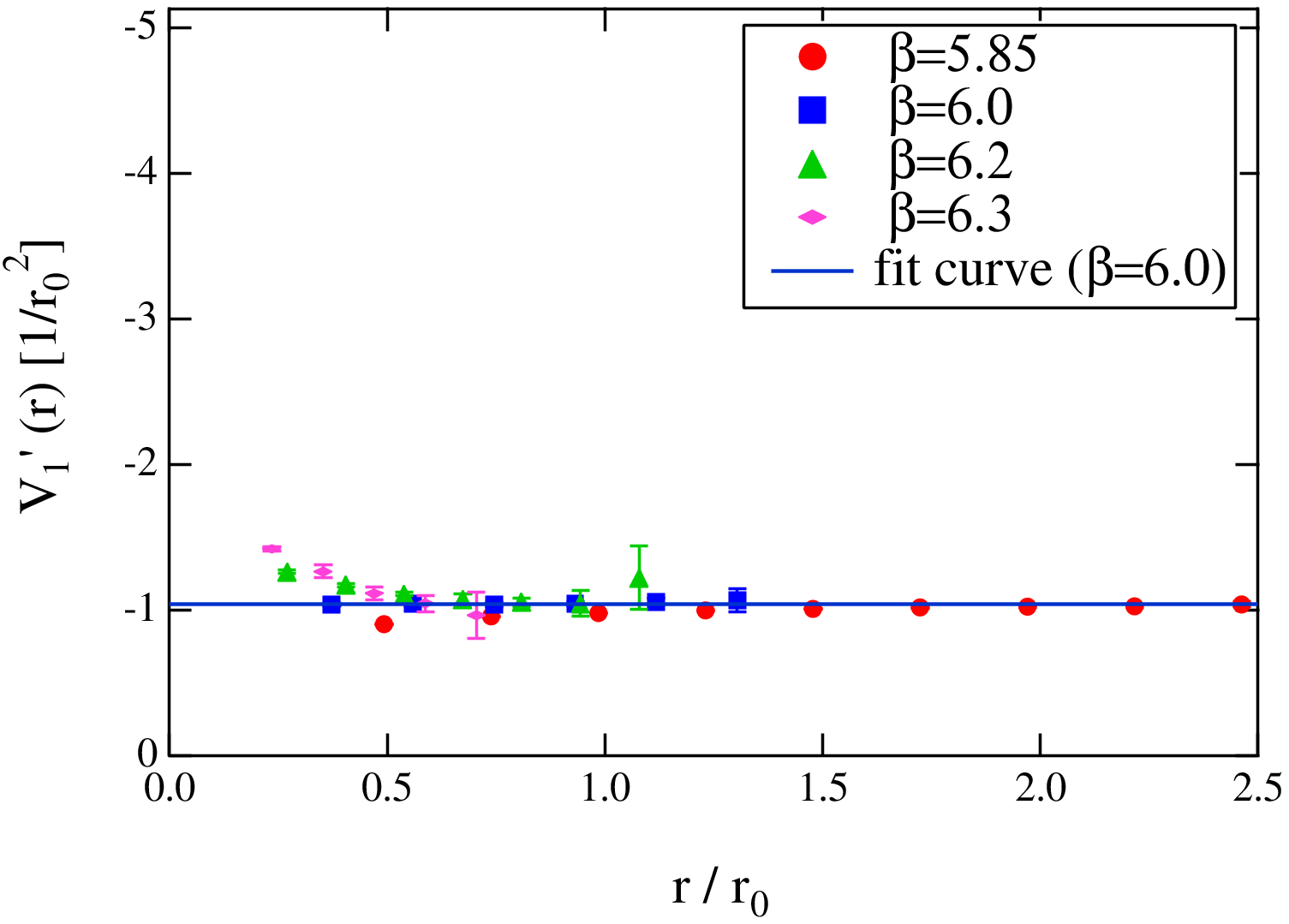}\hspace{0.4cm}\includegraphics[width=\xsize]{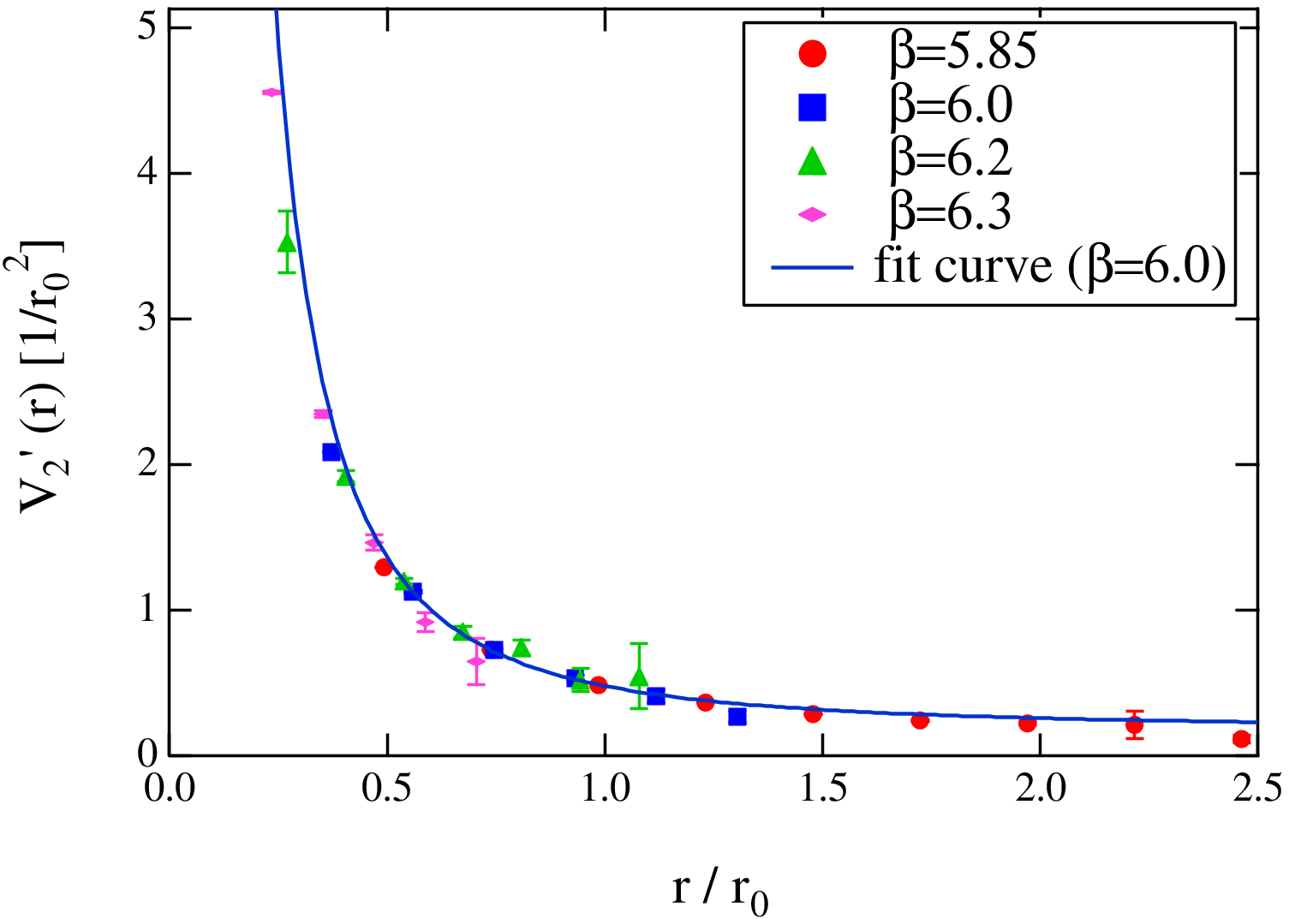}
\caption{Summary of the lattice QCD results used in the present analysis,
the static potential $V^{(0)}$ (upper left), the $O(1/m)$ correction $V^{(1)}$ (upper right),
and the $O(1/m^{2})$ spin-orbit corrections $V_{1}'$ (lower left) and  $V_{2}'$ (lower right).
Solid lines are the fitting curves. }
\label{fig:lattice}
\end{figure}

\section{Heavy quarkonium mass spectra in pNRQCD with the lattice QCD input}

\begin{figure}[!t]
\centering
\includegraphics[width=15.5cm]{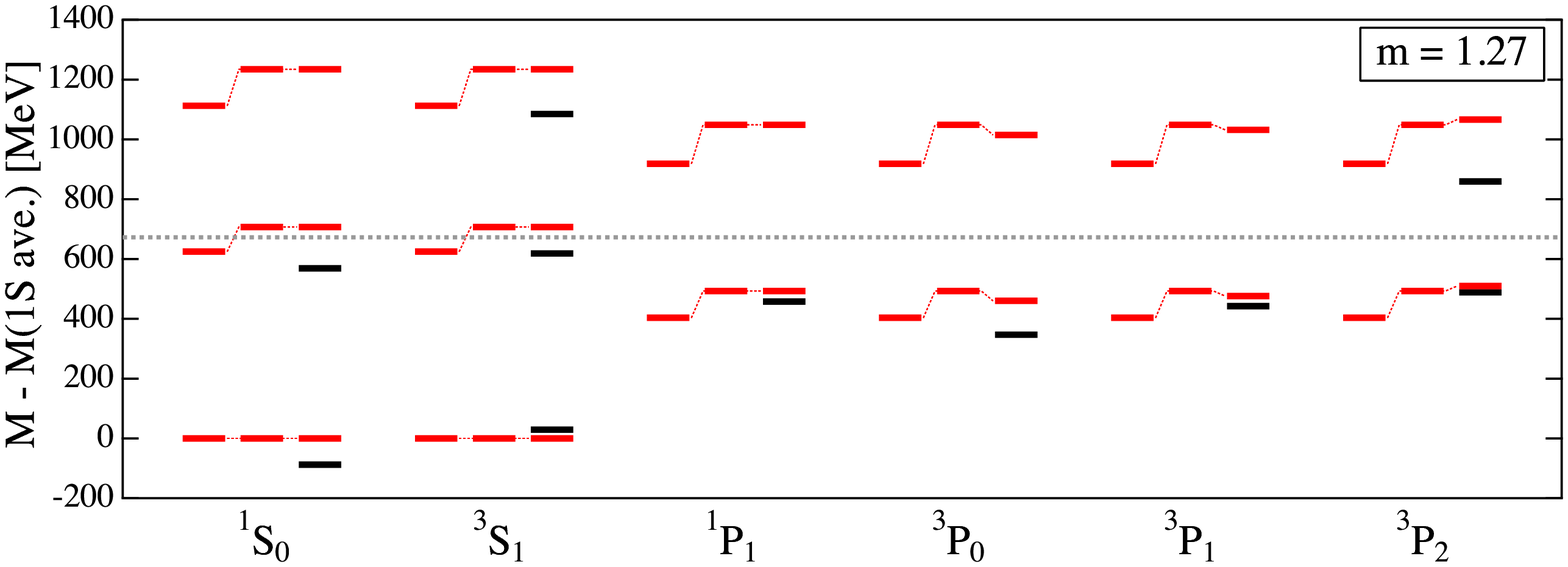}
\caption{The mass spectra of charmonium for various quantum numbers,  ${}^{2S+1}L_{J}$;
a set of three red lines in each state from left to right are the spectra of
$E^{(0)}$, $E^{(0)}+\Delta E^{(1)}$, and $E^{(0)}+\Delta E^{(1)}+\Delta E_{ls}^{(2)}$,
respectively, while the black lines are the experimental data.
The energy levels are normalized at the spin-averaged 1S state.}
\label{fig:spectra-c}~\\
\includegraphics[width=15.5cm]{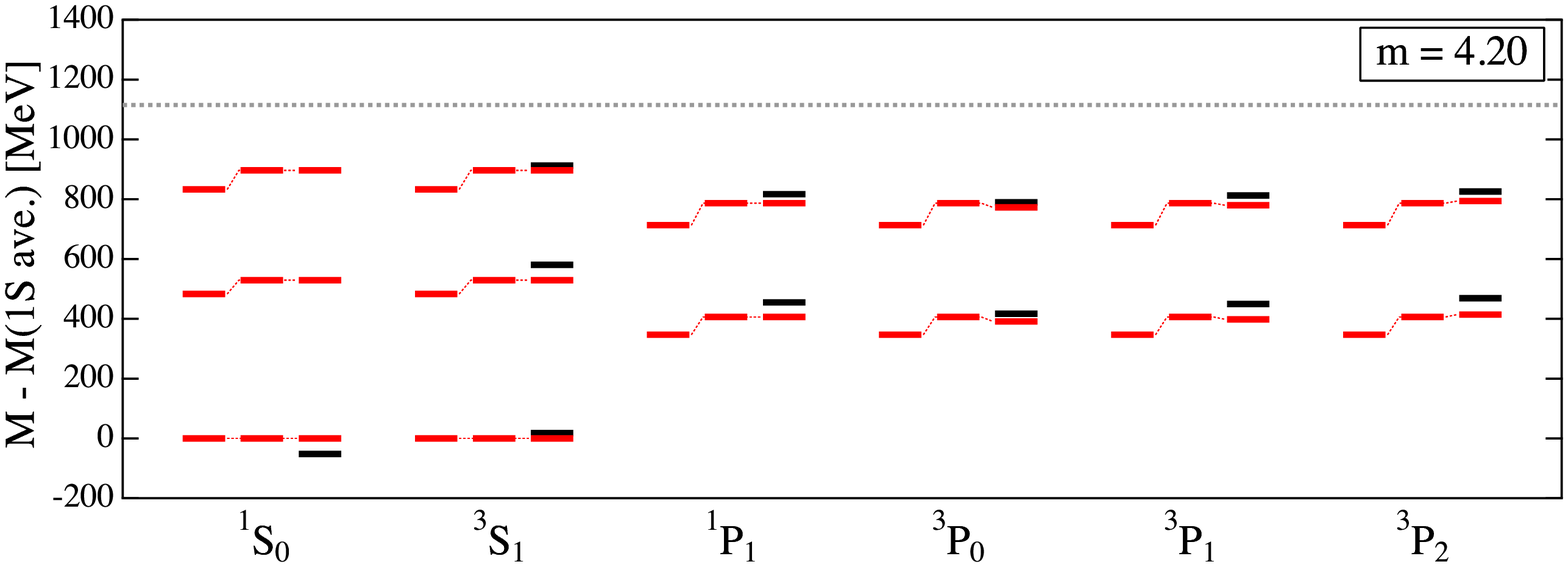}
\caption{The same figure as Fig.~2,
but for bottomonium.}
\label{fig:spectra-b}
\end{figure}

The Schr\"odinger equation 
\be
\left (2m + \frac{\pmb{p}^{2}}{m}+V^{(0)}\right ) \psi^{(0)}=E^{(0)}\psi^{(0)}
\ee
is then solved and  the energy  $\! E^{(0)}\!  $ and the wave function $\! \psi^{(0)}\! $
are computed.
Using the wave function the corrections to $E^{(0)}$ are evaluated 
in the first order perturbation theory,
$\Delta E^{(i)} \!=\!  \langle \psi^{(0)} | V^{(i)} |\psi^{(0)} \rangle /m^{i}$ for $i\ge 1$.
In this way, the heavy quarkonia of various quantum numbers 
are investigated systematically.
The matching coefficients in NRQCD are dependent on the matching scale between
QCD and NRQCD,  and thus different for the charm and bottom quark sector, but 
they are simply assumed here to be the leading ones, $c_{F}=c_{S}=1$, 
which correspond to taking the matching scale to be infinity.
This issue will be discussed in the forthcoming paper.
It should be emphasized that  there is no free parameter
except for the quark masses and a constant shift of the potential.
The quark masses are chosen to be 
$m_{c}=1.27~{\rm GeV}$~(charm) and $m_{b}=4.20~{\rm GeV}$~(bottom) in this analysis.

\par
Figs.~\ref{fig:spectra-c} and~\ref{fig:spectra-b} show the 
mass spectra of charmonium and bottomonium, respectively, 
for various quantum numbers classified by ${}^{2S+1}L_{J}$.
The energy levels are normalized at the spin-averaged 1S state.
A fairly good agreement  can be observed between 
the computed pattern of the mass spectra and the experimental data, especially 
for the states below the $D\bar{D}$ threshold for charmonium and the $B\bar{B}$ threshold for bottomonium,
in which the effect of the $O(1/m)$ correction is remarkable.
As is clear from the $\pmb{l}\!\cdot\!  \pmb{s}$ operator,
the $O(1/m^{2})$ spin-orbit correction affects the splitting among the levels of ${}^{3}P_{J=0,1,2}$ states, 
the fine splitting, where $\langle {}^{3}P_{J}| \pmb{l}\!\cdot\!  \pmb{s}|{}^{3}P_{J}\rangle=-2,-1,1$ for $J=0,1,2$,
respectively.
The splitting is less noticeable
than that caused by the $O(1/m)$ correction as
the matching scale effect of QCD and NRQCD is not yet properly taken into account.
The experimental level order of ${}^{3}P_{J}$ states exhibits
$\Delta E({}^{3}P_{1} -{}^{3}P_{0}) > \Delta E({}^{3}P_{2} -{}^{3}P_{1})$,
but if only the spin-orbit interaction contributes, this will be opposite regardless 
the functional form of the spin-orbit correction due to the factor 
of $\langle {}^{3}P_{J}| \pmb{l}\!\cdot\! \pmb{s}|{}^{3}P_{J}\rangle$.
It would be crucial to take into account the spin-tensor contribution.

\section{Summary}

\par
We have studied the charmonium and bottomonium mass spectra  in
potential nonrelativistic QCD (pNRQCD) with the lattice QCD results  of the heavy quark potential.
In pNRQCD, the heavy quark potential consists of a static potential and relativistic corrections
classified in powers of $1/m$,
and in the present investigation, we have examined 
the effect of the $O(1/m)$ correction and the $O(1/m^{2})$ spin-orbit correction systematically.
We have found that the pattern of the mass spectra is in fairly good agreement with
 the experimental data,
in which the $O(1/m)$ correction gives an important contribution.
This effect has not been taken into account in the conventional phenomenological models.
It is certainly interesting to see all the effect up to the $O(1/m^{2})$ correction in pNRQCD 
on the spectra, which contain the velocity-dependent potentials as well as the other types 
of the spin-dependent corrections.


\begin{thebibliography}{10}

\bibitem{Beringer:1900zz}
J.~Beringer~{\it et al.} (Particle Data~Group), 
{\em  Phys.Rev.} {\bf D86} (2012)  010001.

\bibitem{Caswell:1985ui}
W.~Caswell and G.~Lepage, 
  {\em Phys.Lett.} {\bf B167} (1986)  437.

\bibitem{Bodwin:1994jh}
G.~T. Bodwin, E.~Braaten and G.~P. Lepage, 
  {\em Phys.Rev.}  {\bf D51} (1995) 1125--1171
  [\href{http://xxx.lanl.gov/abs/hep-ph/9407339}{{\tt hep-ph/9407339}}].

\bibitem{Brambilla:2004jw}
N.~Brambilla, A.~Pineda, J.~Soto, and A.~Vairo, 
   {\em Rev.Mod.Phys.} {\bf 77} (2005) 1423
  [\href{http://xxx.lanl.gov/abs/hep-ph/0410047}{{\tt hep-ph/0410047}}].

\bibitem{Brambilla:2000gk}
N.~Brambilla, A.~Pineda, J.~Soto, and A.~Vairo, 
 {\em Phys. Rev.} {\bf D63} (2001) 014023
  [\href{http://xxx.lanl.gov/abs/hep-ph/0002250}{{\tt hep-ph/0002250}}].

\bibitem{Pineda:2000sz}
A.~Pineda and A.~Vairo, 
   {\em Phys.Rev.} {\bf D63}
  (2001) 054007 [\href{http://xxx.lanl.gov/abs/hep-ph/0009145}{{\tt
  hep-ph/0009145}}].

\bibitem{Koma:2005nq}
M.~Koma, Y.~Koma, and H.~Wittig, 
   {\em PoS} {\bf LAT2005} (2006)
  216 [\href{http://xxx.lanl.gov/abs/hep-lat/0510059}{{\tt hep-lat/0510059}}].

\bibitem{Koma:2006si}
Y.~Koma, M.~Koma, and H.~Wittig, 
 {\em Phys.Rev.Lett.} {\bf 97} (2006) 122003
  [\href{http://xxx.lanl.gov/abs/hep-lat/0607009}{{\tt hep-lat/0607009}}].

\bibitem{Koma:2006fw}
Y.~Koma and M.~Koma, 
 {\em  Nucl.Phys.} {\bf B769} (2007) 79--107
  [\href{http://xxx.lanl.gov/abs/hep-lat/0609078}{{\tt hep-lat/0609078}}].

\bibitem{Koma:2009ws}
Y.~Koma and M.~Koma,
  {\em PoS} {\bf LAT2009} (2009) 122
  [\href{http://xxx.lanl.gov/abs/0911.3204}{{\tt arXiv:0911.3204}}].

\bibitem{Koma:2010zz}
Y.~Koma and M.~Koma, 
 {\em   Prog.Theor.Phys.Suppl.} {\bf 186} (2010) 205--210.

\bibitem{Manohar:1997qy}
A.~V. Manohar,
{\em  Phys.Rev.} {\bf D56} (1997) 230--237
  [\href{http://xxx.lanl.gov/abs/hep-ph/9701294}{{\tt hep-ph/9701294}}].

\bibitem{Pineda:1998kj}
A.~Pineda and J.~Soto, 
{\em Phys.Rev.} {\bf D58} (1998) 114011
  [\href{http://xxx.lanl.gov/abs/hep-ph/9802365}{{\tt hep-ph/9802365}}].


\bibitem{Gromes:1984ma}
D.~Gromes, 
  {\em Z.Phys.} {\bf C26} (1984) 401.

\end{thebibliography}

\providecommand{\href}[2]{#2}\begingroup\raggedright\endgroup

\end{document}